\documentclass[reprint,prb,twocolumn,showpacs,superscriptaddress,aps,longbibliography]{revtex4-1}
\usepackage[utf8]{inputenc}

\usepackage{graphicx}
\DeclareGraphicsExtensions{.png,.jpg,.eps,.pdf}

\usepackage{float}
\usepackage{xcolor}
\usepackage{amsmath}
\usepackage{float}
\DeclareMathAlphabet\mathbfcal{OMS}{cmsy}{b}{n}

\begin{document}

\title{Moiré-driven multiferroic order in twisted CrCl$_3$, CrBr$_3$ and CrI$_3$ bilayers}

\author{Adolfo O. Fumega}
\affiliation{Department of Applied Physics, Aalto University, 02150 Espoo, Finland}

\author{Jose L. Lado}
\affiliation{Department of Applied Physics, Aalto University, 02150 Espoo, Finland}

\begin{abstract}
Layered van der Waals materials have risen as a powerful platform to engineer artificial competing states of matter. Here we show the emergence of multiferroic order in twisted chromium trihalide bilayers, an order fully driven by the moiré pattern and absent in aligned multilayers. Using a combination of spin models and ab initio calculations, we show that a spin texture is generated in the moiré supercell of the twisted system as a consequence of the competition between stacking-dependent interlayer magnetic exchange and magnetic anisotropy. An electric polarization arises associated with such a non-collinear magnetic state due to the spin-orbit coupling, leading to the emergence of a local ferroelectric order following the moiré. Among the stochiometric trihalides, our results show that twisted CrBr$_3$ bilayers give rise to the strongest multiferroic order. We further show the emergence of a strong magnetoelectric coupling, which allows the electric generation and control of magnetic skyrmions. Our results put forward twisted chromium trihalide bilayers, and in particular CrBr$_3$ bilayers, as a powerful platform to engineer artificial multiferroic materials and electrically tunable topological magnetic textures.
\end{abstract}

\maketitle

\section{Introduction}

Multiferroic materials display more than one ferroic order at the same time \cite{Hill2000,Spaldin2019,Fiebig2016}, and in particular, they can simultaneously host
magnetic and ferroelectric orders. The existence of multiple symmetry-breaking orders allows having a coupling between electric and magnetic degrees of freedom\cite{Fiebig2005}. 
Over the last two decades, a variety of multiferroic bulk compounds
has been demonstrated\cite{Kimura2003,Hur2004,Gajek2007,Nan2008}, 
providing alternative strategies for 
multifunctional devices\cite{Gajek2007,Pantel2012}.
However, focusing on the realm of two-dimensional (2D) materials,
purely 2D multiferroics have remained elusive until the
recent demonstration of multiferroic order in NiI$_2$ \cite{Fumega2022,Song2022}.
Beyond isolating individual multiferroic monolayers, 
a potential alternative strategy
to realize multiferroic order in van der Waals materials relies  on artificially engineering it from originally
non-multiferroic monolayers\cite{Su2020}.
The electric control of magnetism provided by van der Waals multiferroics
would open radically new ways of
controlling artificial van der Waals matter\cite{Serlin2020,Geisenhof2021,Li2021,Nuckolls2020,superCao2018,Oh2021,Kezilebieke2020,Kezilebieke2022,Park2021,Kim2022,Vao2021,Shen2022,Ruan2021}.

The weak bonding between layers
in van der Waals materials allows combining monolayers in twisted heterostructures. Monolayer 2D materials can display different symmetry-breaking orders\cite{Song2022,NbSe22015,Huang2017,Yuan2019}, constituting a family of minimal building blocks that can be used to artificially engineer other emergent orders.
This strategy has been widely exploited to engineer moiré correlated and topological
states using 2D materials \cite{mottCao2018,superCao2018,Lu2019,Serlin2020,Nuckolls2020,Polshyn2020,PhysRevLett.122.086402,Haavisto2022}.
Recently, this strategy has been extended
to 2D magnetic materials
including chromium trihalides, leading to a variety of twist-induced magnetic orders \cite{Song2021,PhysRevLett.125.247201,Xie2021,Xu2021,PhysRevResearch.3.013027,Akram2021,2021arXiv210309850X,2022arXiv220401636X}. 
However, using twist engineering to realize a multiferroic order has so far remained unexplored.

Here we demonstrate the emergence of a multiferroic state in the family of twisted chromium trihalide CrX$_3$ (X=Cl, Br and I) bilayers
by combining first principles and effective spin Hamiltonians.
We first show the emergence of 
a non-collinear spin texture due to the modulation of the interlayer exchange coupling in the moiré unit cell. Associated with the spin texture an electric polarization emerges as a consequence of spin-orbit coupling (SOC) and the local magnetic non-collinearity in the moiré domains. 
Using ab initio calculations we extract the value of the electric polarization
driven by non-collinear magnetic texture, and demonstrate its dependence
on the halide of CrX$_3$. 
Therefore, we provide a quantification of the resulting multiferroic order.
Furthermore, we analyze the emergent magnetoelectric coupling, 
demonstrating how it allows us to
electrically drive transitions between
different topological spin textures employing an interlayer bias.

\begin{figure*}[ht!]
  \centering
  \includegraphics[width=\textwidth]
        {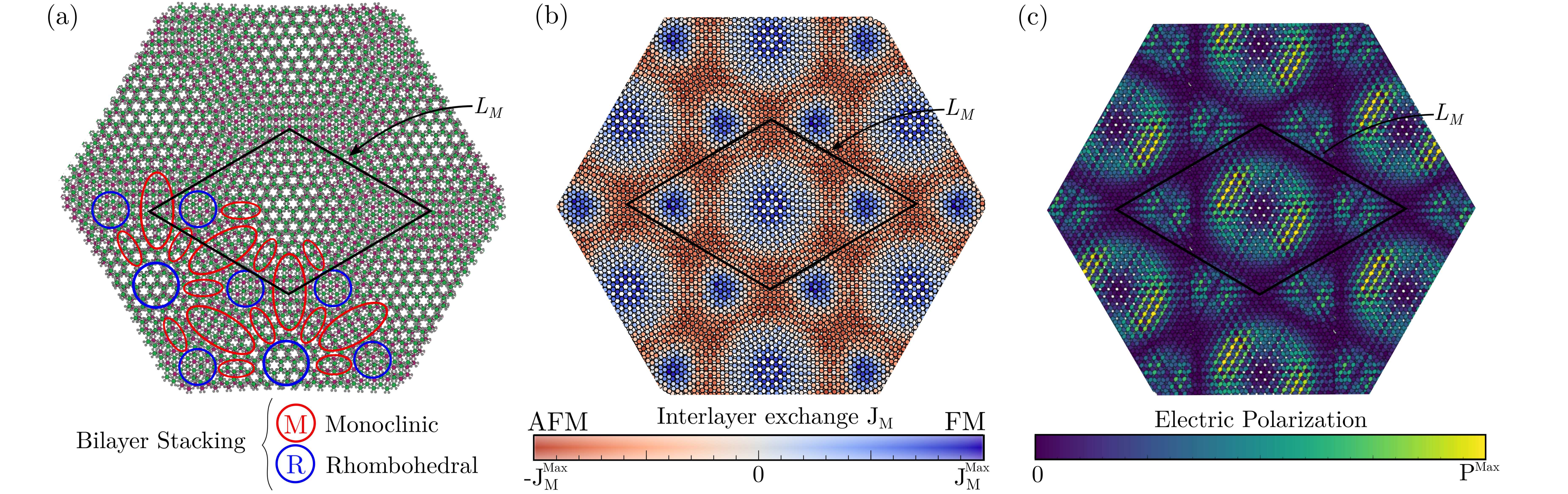}% picture filename
     \caption{(a) Top view of the twisted CrX$_3$ bilayer at small angles. Ligand atoms X are represented in gray, and Cr atoms of the bottom (top) layer in pink (green). The moiré unit cell is described by a lattice parameter of the moiré length scale $L_M$. The emergent moiré pattern leads to regions with monoclinic (M) and rhombohedral (R) stackings, highlighted in red and blue respectively.  (b) Moiré profile for the interlayer magnetic exchange $J_M$ originated from the different stackings. The interlayer magnetic exchange interaction is antiferromagnetic (ferromagnetic) in the monoclinic (rhombohedral) stacking. (c) Module of the local electric polarization associated with the moiré ground state spin texture.}
     \label{Fig:scheme}
\end{figure*}

\section{Results and discussion}

We start our analysis by describing the magnetic order that emerges in twisted CrX$_3$ bilayers. 
The magnetic behavior of CrX$_3$ monolayers can be described by a spin Hamiltonian in a honeycomb lattice

\begin{equation}\label{eq:hamiltonian_intra}
  \mathcal{H}= -\frac{J}{2}\sum_{\langle i,j \rangle }\mathbf{S}_i \cdot \mathbf{S}_j-\frac{A_{v}}{2}\sum_{\langle i,j \rangle}S^z_iS^z_j-A_u\sum_{i}\left(S^z_i\right)^2 + \mathcal{V},
\end{equation}

where $J$ is the first neighbor intralayer ferromagnetic exchange,
taking a value on the order of 2-3 meV\cite{Lado2017,Zhang2015,PhysRevB.106.245111}. $A_{v}$ is the anisotropic magnetic exchange. 
$A_{u}$ is the single-ion anisotropy, which
is dominated by $A_{v}$ in CrI$_3$, CrBr$_3$, and competing with
$A_{v}$ in CrCl$_3$\cite{Tartaglia2020}.
Our analysis will focus on the anisotropy regime leading to
out-of-plane easy axis, that corresponds to CrI$_3$,
CrBr$_3$ and slightly 
strained CrCl$_3$\cite{PhysRevB.98.144411}\footnote{In the absence of strain, CrCl$_3$ shows in-plane anisotropy.}.
The term $\mathcal{V}$ contains additional
contributions that do not qualitatively affect our analysis, including
Kitaev exchange\cite{Xu2018}, biquadratic exchange\cite{Kartsev2020}, 
direct Dzyaloshinskii-Moriya interaction\cite{PhysRevX.8.041028}, 
and dipolar coupling\cite{Lu2020}.

\begin{figure}[h!]
  \centering
  \includegraphics[width=\columnwidth]
        {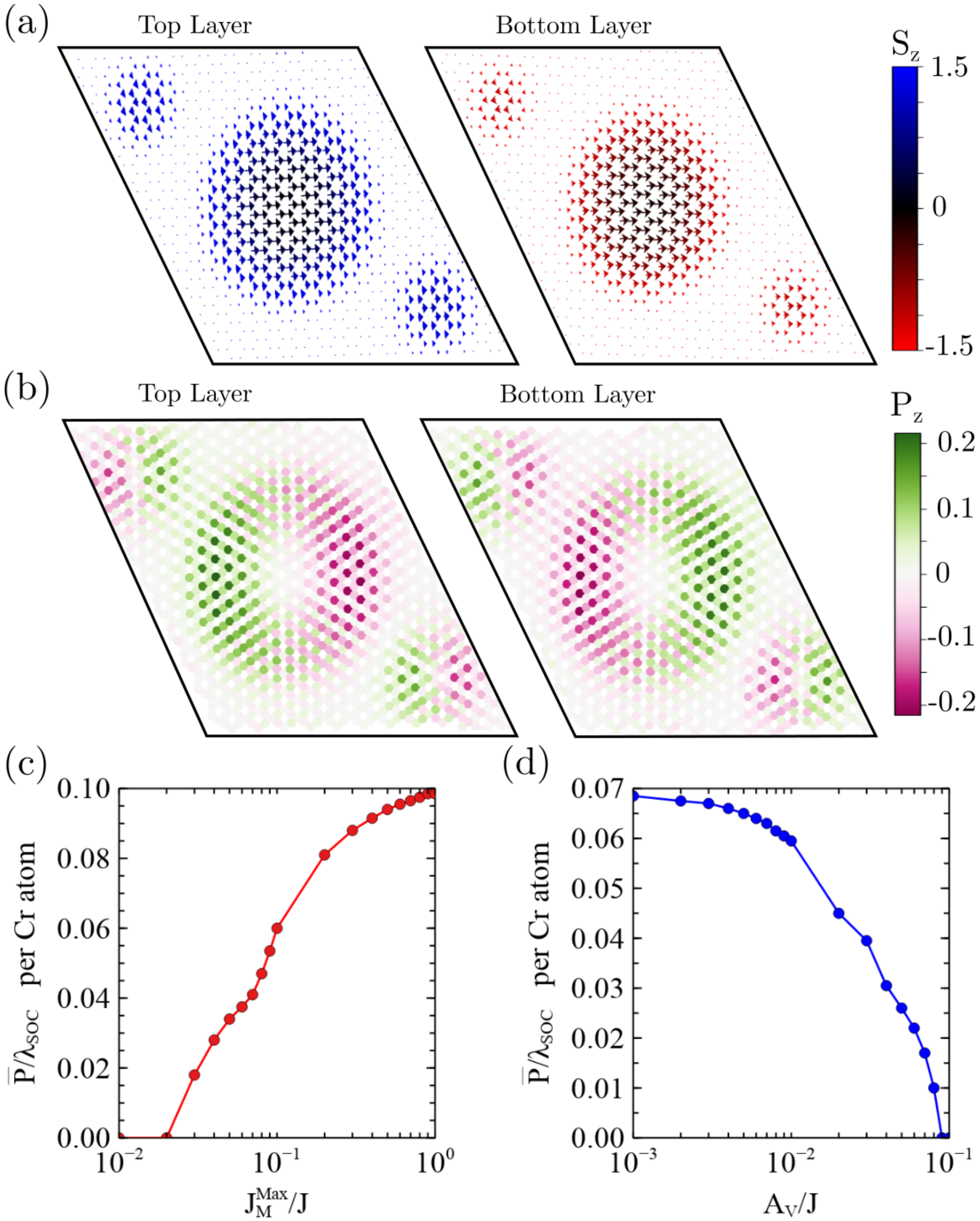}% picture filename
     \caption{Ground state solution of the spin Hamiltonian for the twisted system. (a) Ground state spin texture on the top (bottom) layer is shown in the left (right) panel. $S_x$ and $S_y$ components are depicted as a vector field. The color of the vectors shows the $S_z$ component. (b) Associated electric polarization ($P_z$ component) to the ground state magnetic texture. (c) Electric polarization average of the ground state as a function of the interlayer exchange maximum $J^{max}_{M}$ ($A_{v}/J=0.01$ and $A_{u}/J=0.001$). At low values of $J^{max}_{M}$ no spin texture is formed. (d) Electric polarization average of the ground state as a function of the anisotropic magnetic exchange $A_{v}$ ($J^{max}_{M}/J=0.1$ and $A_{u}/J=0.001$). At high values of $A_{v}$ no spin texture is formed.}
     \label{Fig:spinH}
\end{figure}

Considering now two layers of CrX$_3$ stacked together an interlayer magnetic exchange $J_M$ will arise. A moiré pattern like the one shown in Fig. \ref{Fig:scheme}a emerges for twist angles lower than $3^{\circ}$. Depending on the stacking between layers two different regions can be distinguished, monoclinic and rhombohedral. 
Associated with these different regions the sign of $J_M$ will change, i.e., $J_M$ is ferromagnetic (positive) in the rhombohedric stacking and antiferromagnetic (negative) in the monoclinic one\cite{Sivadas2018,Soriano2019,Chen2019,Gibertini2020}. This leads to a modulation of $J_M$ in the moiré supercell like the one shown in Fig. \ref{Fig:scheme}b\cite{Song2021,PhysRevLett.125.247201,Xie2021,Xu2021,Soriano2022}\footnote{In the case of CrCl$_3$ the stacking dependence of the interlayer magnetic exchange has not been fully established. Ab initio calculations have found a highly functional dependence on the interlayer magnetic exchange\cite{Akram2021,Klein2019}. However, it has also been shown, that this interlayer coupling is weak and external fields, can drive bilayer CrCl$_3$ to the same scenario as the other two halides \cite{Klein2019}.}. Therefore, to model the twisted CrX$_3$ bilayer we add the interlayer interaction to the Hamiltonian of Eq. (\ref{eq:hamiltonian_intra}) as

\begin{equation}\label{eq:hamiltonian_inter}
    \mathcal{H}_{Inter}= -\frac{1}{2}\sum_{i,j}J_{M}(\mathbf{r}_i,\mathbf{r}_j)
    \mathbf{S}_i \cdot \mathbf{S}_j,
\end{equation}

where $J_{M}(\mathbf{r}_i,\mathbf{r}_j)$ is the site-dependent interlayer exchange \footnote{A detailed explanation of the interlayer exchange parametrization and the robustness of the results against perturbations to it, such as the twist angle, are included in the Supplemental Material.}. Since CrX$_3$ is composed of Cr$^{3+}$ with a spin state S=$3/2$, we can solve in a classical way the spin Hamiltonian for the twisted system \footnote{Computational details about the minimization procedure to obtain the ground state are included in the Supplemental Material}. The ground state magnetic order is depicted in Fig. \ref{Fig:spinH}a. We can see that a non-collinear magnetic texture emerges between ferromagnetic and antiferromagnetic regions in agreement with previous theoretical\cite{PhysRevResearch.3.013027,Akram2021} and experimental results\cite{Song2021,PhysRevLett.125.247201,Xie2021,Xu2021,2021arXiv210309850X,2022arXiv220401636X}. Taking this as the starting point, we now show that in the presence of spin-orbit coupling, this topologically-trivial spin texture leads to the emergence of an electric polarization $\mathbf{P}_{ij}$ between first-neighbor spins in the same layer separated by a distance $\mathbf{r}_{ij}$ due to the inverse Dzyaloshinskii-Moriya (DM) mechanism\cite{PhysRevLett.95.057205, PhysRevLett.96.067601}

\begin{equation}\label{eq:electric_polarization}
    \mathbf{P}_{ij}=\alpha\lambda_{SOC}\left(\mathbf{r}_{ij}\times\left(\mathbf{S}_i\times\mathbf{S}_j\right)\right),
\end{equation}

where $\lambda_{SOC}$ is a coefficient that controls the strength of the spin-orbit coupling and $\alpha$ is a proportionality constant that depends on the electronic structure and crystal environment, which is similar for the three chromium trihalides. The polarization emerges at the middle point between the two neighboring spins $\mathbf{S}_i$ and $\mathbf{S}_j$. From Eq. (\ref{eq:electric_polarization}) we can clearly see the requirement of non-collinearity and the presence of SOC to produce an electric polarization. For the emerging ground state spin texture of the twisted system (Fig. \ref{Fig:spinH}a), the associated electric polarization when SOC is introduced is shown in Fig. \ref{Fig:spinH}b, with $P_z$ the dominant component. 
From Figs. \ref{Fig:spinH}ab, we can observe that an electric polarization emerges in both layers in the areas where the non-collinear spin texture occurs. The polarization emerges locally, and for the ground state spin texture the net electric polarization is zero. Therefore, to analyze the strength of the multiferroic order as a function of the parameters that appear in the spin Hamiltonian, we will consider an average of the electric polarization module $\overline{P}$. The module of the electric polarization in the moiré system associated with the ground state spin texture can be seen in Fig. \ref{Fig:scheme}c. We can observe there, that the ground state breaks the C$_6$ symmetry of the original system leading to a C$_2$ symmetry. The lifted C$_6$ symmetry stems from a spontaneous symmetry breaking of the ground state due to the ratio between the parameters entering the spin Hamiltonian, analogous to the symmetry breaking associate to stripy magnetic orders. Figure  \ref{Fig:spinH}c shows the average polarization as a function of the maximum interlayer exchange $J^{max}_{M}$. At low values, the twisted system behaves like two independent layers, remaining ferromagnetic, no spin texture emerges and consequently, there is no electric polarization. By increasing the interlayer coupling, non-collinearity appears and with it an associated electric polarization in virtue of Eq. (\ref{eq:electric_polarization}). 
Figure \ref{Fig:spinH}d shows the average polarization as a function of the anisotropic exchange $A_{v}$. At high values, the twisted system tends to align the magnetic moments out of the plane. Therefore, no spin texture occurs and thus there is no electric polarization. This situation happens for $A_{v}/J>0.1$, so CrI$_3$ would be on the verge of displaying a multiferroic behavior due to its strong uniaxial anisotropy \cite{Tartaglia2020}.

\begin{figure}[h!]
  \centering
  \includegraphics[width=\columnwidth]
        {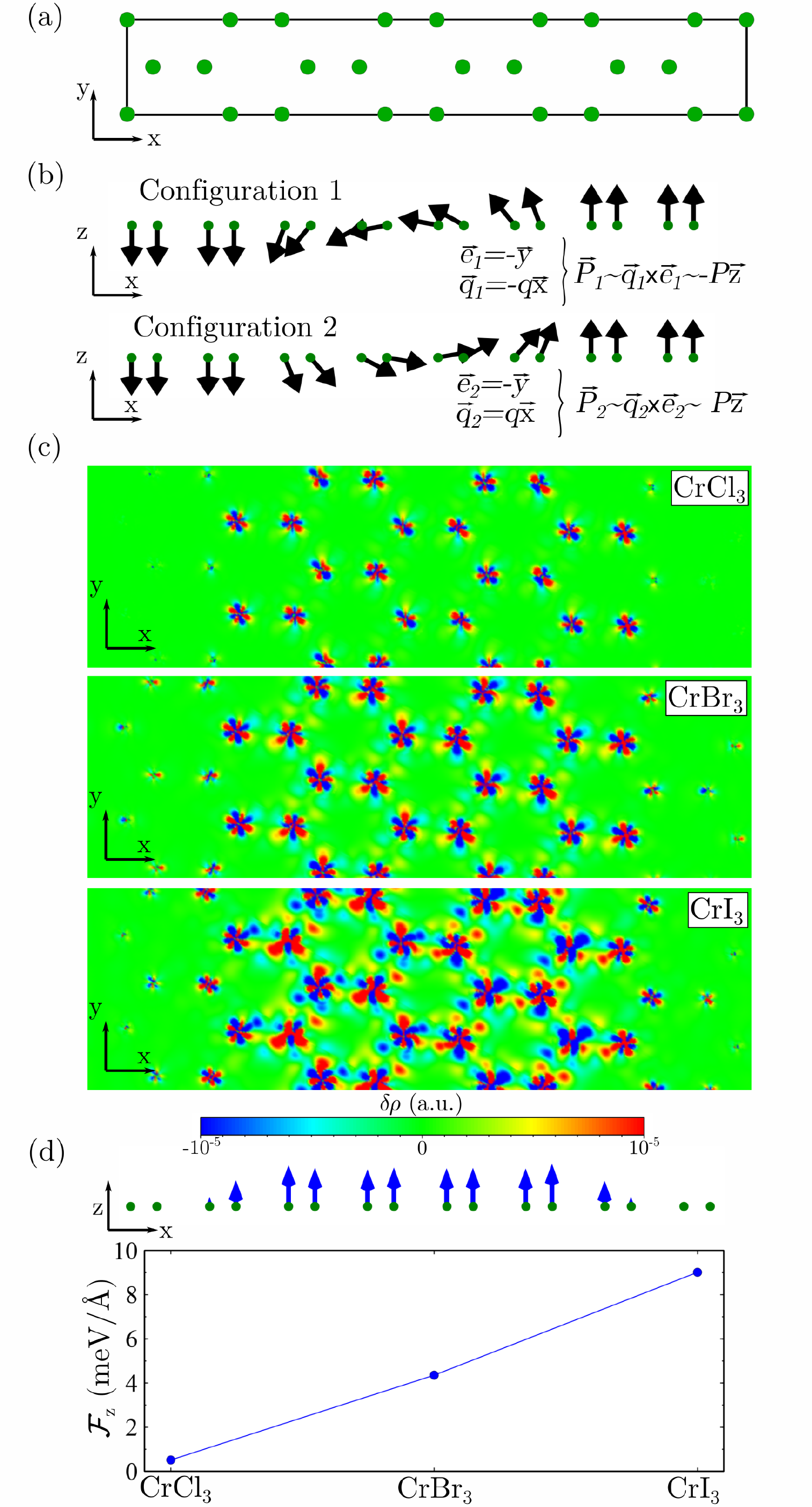}% picture filename
     \caption{(a) Unit cell used in the ab initio calculations, Cr atoms depicted in green, halide atoms omitted for clarity. (b) Equivalent non-collinear magnetic configurations with opposite spin propagation vector $\mathbf{q}$ and same helicity $\mathbf{e}$, and hence opposite electric polarization. (c) Electronic density difference  $\delta\rho$ between both equivalent configurations in panel (b) for the three trihalides in the Cr-atoms plane. (d) Force difference between the configurations of the panel (b). A net ferroelectric force emerges in the z-direction in the Cr atoms where the magnetism is non-collinear. An average of the ferroelectric force for each trihalide shows the dependence of the electric polarization with the ligand.}
     \label{Fig:dft}
\end{figure}

We now demonstrate and quantify, using density functional theory  calculations\cite{HK},
the emergent electric polarization that we have seen that appears in the non-collinear moiré system in virtue of eq. (\ref{eq:electric_polarization}). 
Performing first-principles calculations in a full twisted CrX$_3$ bilayer with spin-orbit coupling and non-collinear magnetism is well beyond
the current computational capabilities.
However, since the electric polarization arises
locally, the ab initio analysis can be performed in a 
system like the one shown in Fig. \ref{Fig:dft}a by imposing a magnetic texture in the CrX$_3$
layer like the one shown in Fig. \ref{Fig:dft}b, that resembles the kind of spin texture found in the ground state between rhombohedric and monoclinic regions. We set the same out-of-plane spin texture to the three compounds to systematically extract the effect of the halide atom and quantification of the inverse DM interaction \footnote{An in-plane spin texture provides results on the same order of magnitude for CrCl$_3$.}.  
The associated polarization of a non-collinear texture given in eq. (\ref{eq:electric_polarization}) can be rewritten in terms of a spin spiral propagation vector  $\mathbf{q}$ and the spin rotation axis $\mathbf e = (0,-1,0)$, leading to an electric polarization of the spin texture\cite{PhysRevLett.95.057205, PhysRevLett.96.067601}

\begin{equation}\label{eq:electric_pol_spin_spiral}
    \mathbf P =\beta \lambda_{SOC} (\mathbf q \times \mathbf e),
\end{equation}

where $\beta$ is a proportionality constant that depends on the electronic structure and crystal environment and are similar for the three compounds.
To demonstrate and quantify the emergence of an electric polarization 
we performed \emph{ab initio} calculations in two equivalent magnetic configurations (Fig. \ref{Fig:dft}b) with the same spin rotation vector $\mathbf{e}$, but with opposite spin propagation vector $\mathbf{q}$. Therefore, an opposite electric polarization will emerge in each of the configurations. The emergence of the electric polarization in the spin texture is directly obtained by taking the difference between the two equivalent configurations. 
This procedure provides a direct methodology to extract
electric polarization stemming from non-collinear magnetic order \footnote{A detailed explanation about the DFT calculations can be found in the Supplemental Material.}. Modifying the $q$-vector of
the spiral, i.e. the size of the supercell, will change the total value of the ferroelectric polarization obtained since it will modify the non-collinear spin order controlling the inverse DM interaction.
In our analysis, we use the same spin texture for every chromium
trihalide, and hence we can extract the contribution coming
from the spin-orbit coupling to the inverse DM interaction.

The emergence of an electric polarization is accompanied by a reconstruction of the electronic density $\rho$. Therefore, we can analyze the difference in the electronic density $\delta\rho$ between both configurations. This is shown in  Fig. \ref{Fig:dft}c for the three chromium 
halides. We observe that the electronic reconstruction increases by taking a heavier halide.
Thus, for the same spin texture, we can see that CrI$_3$ will produce
the strongest ferroelectric polarization. 
This result is a consequence of the increase of the spin-orbit coupling when one goes down in the halide group, thus demonstrating the effective equations (\ref{eq:electric_polarization}) and (\ref{eq:electric_pol_spin_spiral}) governed by the SOC prefactor $\lambda_{SOC}$.

The reconstruction of the electronic density will lead to the appearance of a ferroelectric force in the atoms in the spin texture in the direction of the emergent electric polarization\cite{Fumega2022}. 
Therefore, we can compute the force difference between both configurations in Fig. \ref{Fig:dft}b to provide a direct quantification of the electric polarization in CrX$_3$. 
Figure \ref{Fig:dft}d shows the force difference between both configurations. We can see that the forces emerge only 
in the Cr atoms in which the non-collinear magnetism is present. Moreover, we can see that the forces emerge in the z-direction, indicating the direction of the electric dipole, and as expected from the schematic in Fig. \ref{Fig:dft}b. 
We can quantify the dependence on the halide by taking the average of the force module among Cr-atoms for each of the compounds, as shown in \ref{Fig:dft}d. 
Taking a heavier halide produces an increase in the ferroelectric force, as expected from the increase of the spin-orbit coupling. As a reference, NiI$_2$, a 2D multiferroic governed by this same mechanism of spin-orbit coupling and non-collinear magnetism, leads to ferroelectric forces of $\approx 20$ meV/Å \cite{Fumega2022}. Therefore, this confirms that spin textures produced in twisted CrX$_3$ bilayers lead to the emergence of an electric polarization with strong magnetoelectric coupling. 

We now elaborate on some conclusions on the strength of multiferroic order in twisted CrX$_3$ bilayers
considering together the results obtained from the low energy model and the ab initio calculations that allowed us to provide a quantification of the inverse DM interaction in these magnetic moiré systems \footnote{See the Supplemental material for a more detailed description of how these conclusions are derived.}.
On the one hand, our ab initio DFT methods
show that as the halide becomes heavier, 
the ferroelectric polarization becomes stronger.
On the other hand, the low energy model shows that the anisotropic exchange has a detrimental impact on the formation of the spin texture and hence on the multiferroic order.
In particular, in CrI$_3$ the strong anisotropic exchange 
might partially quench the formation of a sizable 
non-collinear magnetic texture.
In CrCl$_3$, the small SOC would yield a comparably weak 
ferroelectric polarization despite the formation of the non-collinear texture.
Therefore, the optimal twisted bilayer that yields the strongest ferroelectric
polarization would be CrBr$_3$,
or ultimately, a bilayer of an intermediate stoichiometry between CrBr$_3$ and CrI$_3$\cite{Tartaglia2020}.

\begin{figure}[h!]
  \centering
  \includegraphics[width=\columnwidth]
        {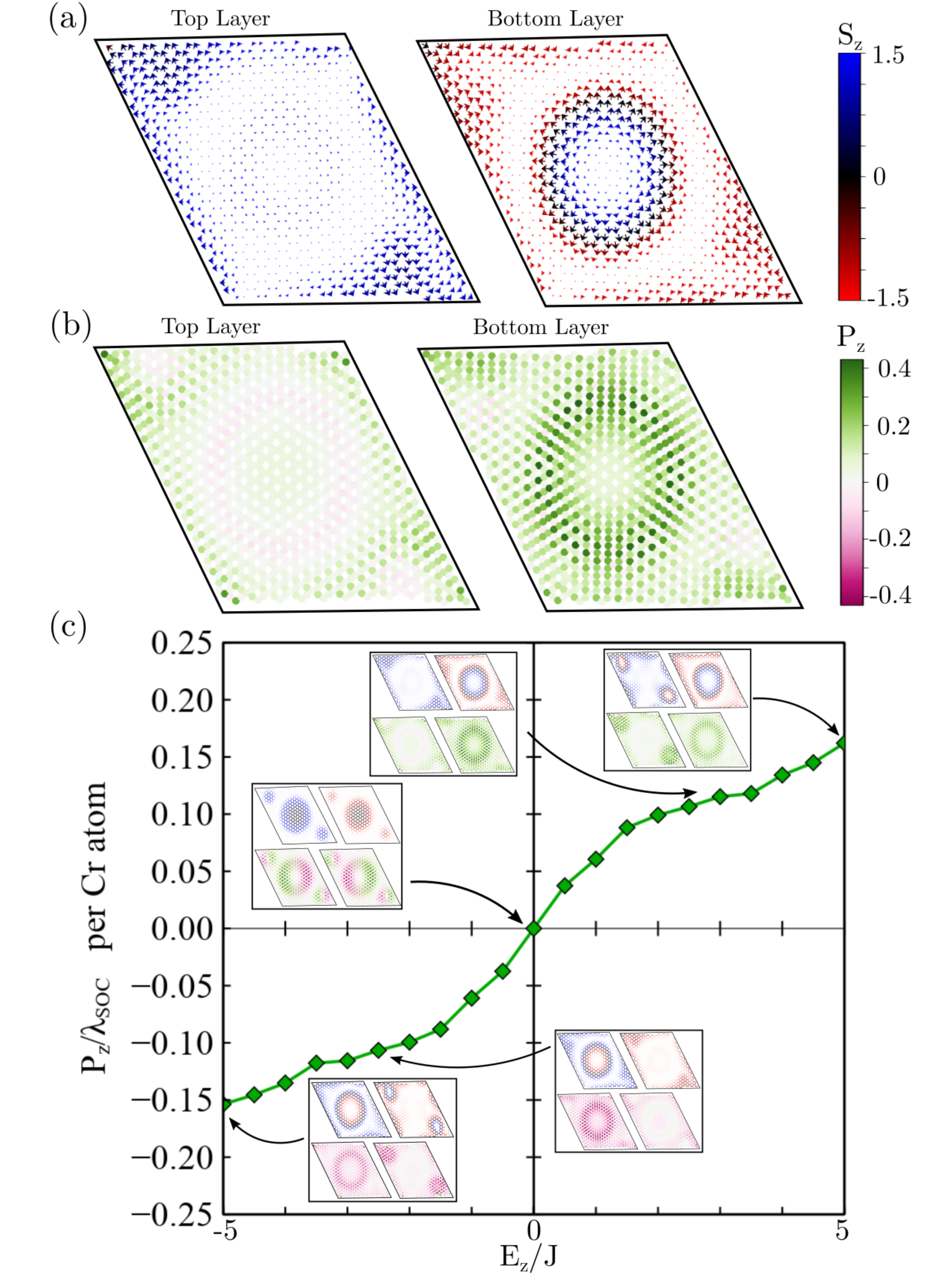}% picture filename
     \caption{(a) Spin texture in moiré unit cell at $E_z/J=3$, a magnetic skyrmion emerges in the bottom layer. (b) Associated electric polarization to the spin texture. (c) Evolution of the net electric polarization in the z-direction $P_z$ as a function of the electric field. Different spin textures appear as a function of the electric field: ground state spin texture at  $E_z/J=0$, a magnetic skyrmion appears in the AA stacking area in one of the layers at $E_z/J=\pm3$, and a second pair of magnetic skyrmions emerge in the other layer around the AB/BA stacking regions at  $E_z/J=\pm5$.}
     \label{Fig:Magnetoelectric}
\end{figure}

Finally, we analyze the magnetoelectric coupling present in twisted CrX$_3$ bilayers \footnote{Importantly, to drive conclusions about the strength of the magnetoelectric coupling we are considering the parameters corresponding to CrBr$_3$, which is found to display the strongest multiferroic order.}. To do so, we now include in the low energy Hamiltonian a coupling to an external electric field $\mathbf{E}=(0,0,E_z)$  perpendicular to the twisted system in the $z$-direction 

\begin{equation}\label{eq:electric_field_coupling}
     \mathcal{H}_{E}=\frac{1}{2}\sum_{ij}\mathbf E \cdot \mathbf P_{ij}.
\end{equation}

We now show how this interlayer bias allows controlling the magnetic order due to the
emergent multiferroicity\cite{2022arXiv220403837C}. 
Figures \ref{Fig:Magnetoelectric}a and \ref{Fig:Magnetoelectric}b show the spin texture and the associated electric polarization at $E_z/J=3$. We can observe that the ground state spin texture shown in Fig. \ref{Fig:spinH}a gets significantly modified due to the strong magnetoelectric coupling, leading to the formation of a magnetic skyrmion around the AA rhombohedral stacking in one of the layers\cite{Tong2018,PhysRevB.104.014410,PhysRevB.104.L100406,PhysRevResearch.3.013027,Akram2021,Hejazi2020,Wu2022,Fawei2023,2022arXiv220605264K}.
Interestingly, such a magnetic state features
a non-zero total electric polarization in the $z$-direction. 
The evolution of the total electric polarization in the moiré supercell in the $z$-direction
as a function of the external electric field is shown in Fig. \ref{Fig:Magnetoelectric}c. The different bumps in $P_z$ as a function of the electric field indicate transitions to different non-trivial spin textures which are shown as insets in Fig. \ref{Fig:Magnetoelectric}c. Starting at $E_z/J=0$ in the ground state spin texture, at  $E_z/J=\pm3$ the magnetic skyrmion shown in Fig. \ref{Fig:Magnetoelectric}ab arises. At $E_z/J=\pm 5$ a second pair of magnetic skyrmions emerge in the other layer around the AB/BA stacking regions \footnote{See Supplemental material for a discussion on topological magnetoelectric couplings associated to skyrmions.}. The critical electric field required to produce a transition to the skyrmionic phase is inversely proportional to the strength of the $\lambda_{SOC}$.
As a reference, $J\approx 3$ meV and $\alpha\lambda_{SOC}\approx 10^{-4}$e (e electron charge) in twisted CrBr$_3$ bilayers, implying that such magnetic transitions
can be experimentally driven via gating
at voltages of $1-10$ V\cite{2022arXiv220403837C}. 
This brings twisted CrBr$_3$ bilayers as a promising platform for the electric control of non-trivial magnetic textures, and ultimately
as a platform for magnetoelectric skyrmionics.

\section{Conclusions}

To summarize, we have demonstrated that twisted CrX$_3$ bilayers develop a multiferroic order
due to the interplay between the moiré structure, non-collinear magnetic order, and spin-orbit coupling. We have provided a quantification of the strength of the multiferroic order for this family of compounds, finding that, among the stoichiometric chromium trihalides, CrBr$_3$ is expected to display the strongest multiferroic and magnetoelectric coupling. Furthermore, we have shown that this strong magnetoelectric coupling allows an electrical control of non-trivial magnetic textures 
using an interlayer bias. Our results put forward a strategy to design a new family of artificial multiferroics with a strong magnetoelectric coupling based on twisted
magnetic van der Waals materials.

\section*{Data availability statement}

All data that support the findings of this study are included within the article (and any supplementary files).

\section*{Acknowledgements}

We acknowledge the computational resources provided by
the Aalto Science-IT project,
and the financial support from the
Academy of Finland Projects No. 331342, No. 336243 and No. 349696
and the Jane and Aatos Erkko Foundation. We thank P. Liljeroth for useful discussions.

\end{document}